# Odd-Parity Selection in Parity-Forbidden Electronic Transitions Revealed by Mn$^{4+}$ Fluorescence Spectroscopy

Yitong Wang[1], Fei Tang[2], Jiqiang Ning[1], and Shijie Xu[1*]

[1]Department of Optical Science and Engineering, College of Future Information Technology, Fudan University, 2005 Songhu Road, Shanghai 200438, China

[2]Jiangsu Key Laboratory of Advanced Laser Materials and Devices, School of Physics and Electronic Engineering, Jiangsu Normal University, Xuzhou 221116, China

[*]Corresponding author, email: xusj@fudan.edu.cn

Abstract: Mn$^{4+}$-doped fluoride phosphors represent a significant class of narrow band red-emitting materials, whose luminescent properties are profoundly influenced by electron-phonon coupling. However, the parity-forbidden nature of these electronic transition systems is incompatible with the conventional Condon approximation, which is widely adopted in the classic theories such as the Huang-Rhys theory, a framework established on the assumption of parity-allowed electric dipole transitions. This results in a critical knowledge gap regarding the principles governing the phonon sidebands of parity-forbidden electronic transitions. This study experimentally reveals a pronounced parity-dependent intensity distribution in the phonon sidebands of these systems: significantly suppressed even-order sidebands and normally observed odd-order sidebands. To elucidate the phenomenon, we extend the Huang-Rhys theory to parity-forbidden systems by incorporating the Herzberg-Teller approximation into the treatment of the transition matrix elements. The improved theory successfully uncovers the physical mechanism behind the strong suppression of the even-order sidebands in the parity-forbidden systems, in which the Huang-Rhys factor is derived as $S = \sqrt{2I_3/9I_1}$. This work not only reveals new findings regarding the phonon sidebands of the parity-forbidden electronic transition systems, but also establishes an improved theoretical framework for understanding the electron-phonon coupling mechanisms of color centers in solids.

## 1 Introduction

Solid-state luminescent materials are pivotal for modern lighting, display, and detection technologies.[1-3] In recent years, a significant class of red-emitting phosphors of Mn$^{4+}$-doped fluoride phosphors (e.g., K$_2$SiF$_6$:Mn$^{4+}$) have garnered considerable attention due to their exceptional luminescent properties, particularly their high color purity and high quantum efficiency.[4-9] Typically, $Mn^{4+}$ ions occupy the octahedral center of $[MF_6]^{2-}$ (M = Si, Ge, Ti, etc.) crystal sites, with their red emission originating from the spin- and parity-forbidden $^2E_g \rightarrow {}^4A_{2g}$ d-d transition.[10,11] It has been widely recognized that the fluorescence spectrum of the $K_2SiF_6$:$Mn^{4+}$ phosphors is characterized by the extremely weak zero-phonon line (ZPL) and its strong 1st-order phonon sidebands of several quasi-localized vibrational modes (i.e., $v_3$, $v_4$, and $v_6$ modes of $[MnF_6]^{2-}$ octahedral coordination). No higher-order phonon sidebands have yet been demonstrated so far. Such spectral characteristics are severally inconsistent with the well-established understanding based on classic theoretical models such as Huang-Rhys theory. In this Letter, a critical yet unconventional spectral characteristic was firmly unraveled in the phonon sidebands of the $K_2SiF_6$:$Mn^{4+}$ and $Na_2SiF_6$:$Mn^{4+}$ phosphors: The 3rd- and even 5th-order phonon sidebands were identified in the high-resolution fluorescence spectra, whereas the 2nd- and 4th- etc. even-order ones were found to be significantly suppressed. These findings pose a challenge to the existing theoretical frameworks for fluorescence of electron-phonon coupling systems.

It is well known that the classical Huang-Rhys theory has served as the fundamental framework for understanding phonon sideband characteristics.[12-15] Based on the assumption of parity-allowed electric dipole transitions, this theory employs the Condon approximation for the electronic transition matrix element and derives a dimensionless constant, widely known as the Huang-Rhys factor $S$, to quantitatively characterize the electron-phonon coupling strength. It successfully predicts and explains the spectral structure featuring multiple phonon sidebands flanking the ZPL symmetrically, with their peak intensities approximately obeying a Poisson distribution as a function of phonon order.[13-15]

Nevertheless, the classic theoretical treatments and conclusions encounter serious difficulties in explaining the fluorescence spectra of the $Mn^{4+}$-activated phosphors. The fundamental incompatibility may originate from the electronic transition nature of this material system: The spin- and parity-forbidden nature of the $^2E_g \rightarrow {}^4A_{2g}$ transition causes severely-suppressed ZPL intensity.[4,5,11] This transition nature basically contradicts the foundational assumption of parity-allowed electric dipole transitions under the Condon approximation in the classic theories such as Huang-Rhys theory.[12] Consequently, the traditional theoretical frameworks are inapplicable to the phonon-assisted luminescence in the $Mn^{4+}$-activated phosphors with strong parity-forbidden electronic transitions. This inherent contradiction not only leaves a void in theoretical interpretation of the observed parity-dependent phonon-assisted luminescence phenomenon, but also necessitates a reconsideration of the Condon approximation widely adopted in the classic theories.[16]

To elucidate the observed unique odd-parity phenomenon, we modify the classic theory by incorporating Herzberg-Teller approximation in evaluating the transition matrix elements, thereby extending it to the parity-forbidden electronic transition systems. New theoretical derivations successfully uncover the physical origin of parity-dependent phonon sideband intensities in such material systems: Parity selection rules suppress the 0th-order term of the transition matrix, resulting in significant attenuation of even-order phonon sidebands with overall odd parity. The characteristic spectral intensity distribution observed in the $Mn^{4+}$-activated fluoride phosphors, i.e., notably attenuated 2nd-order sidebands versus clearly observable 1st-, 3rd-, and even 5th-order features, provides the first experimental validation of the improved theoretical prediction.

## 2  Experimental and Results

The $K_2SiF_6$:$Mn^{4+}$ (KSF) and $Na_2SiF_6$:$Mn^{4+}$ (NSF) phosphor samples employed in this study were synthesized via a two-step chemical co-precipitation method and subsequently characterized by X-ray diffraction.[17] Their micro-photoluminescence and Raman spectra were acquired using a home-assembled multi-function integrated micro-spectroscopic system. This system employs a Horiba iHR550 monochromator with a spectral resolution of 0.005 nm, and achieves precise temperature control from 5 K to 300 K via a helium-closed cycle cryostat (Montana Instruments, S100). The photoluminescence (PL) spectra of the studied phosphors were optically excited using a 405 nm CW laser, while Raman scattering signals were excited with a 532 nm solid laser.

Figure 1a displays the Raman spectrum of the KSF sample at 8 K, while Figure 1b illustrates its corresponding crystal structure. The KSF phosphor crystallizes in cubic lattice, where $Mn^{4+}$ ions substitute $Si^{4+}$ sites, forming $[MnF_6]^{2-}$ octahedral coordination with six surrounding $F^-$ ions.[5,11,17,18] Under $O_h$ point group symmetry, this octahedron exhibits six fundamental vibrational modes denoted as $v_1$-$v_6$.[19] The Raman spectrum clearly resolves three characteristic vibration peaks ($v_1$, $v_2$, $v_5$) assigned to the $[SiF_6]^{2-}$ octahedra and three additional peaks ($v_1$, $v_2$, $v_5$) to $[MnF_6]^{2-}$ octahedra, all explicitly labeled in the spectrum.[20]

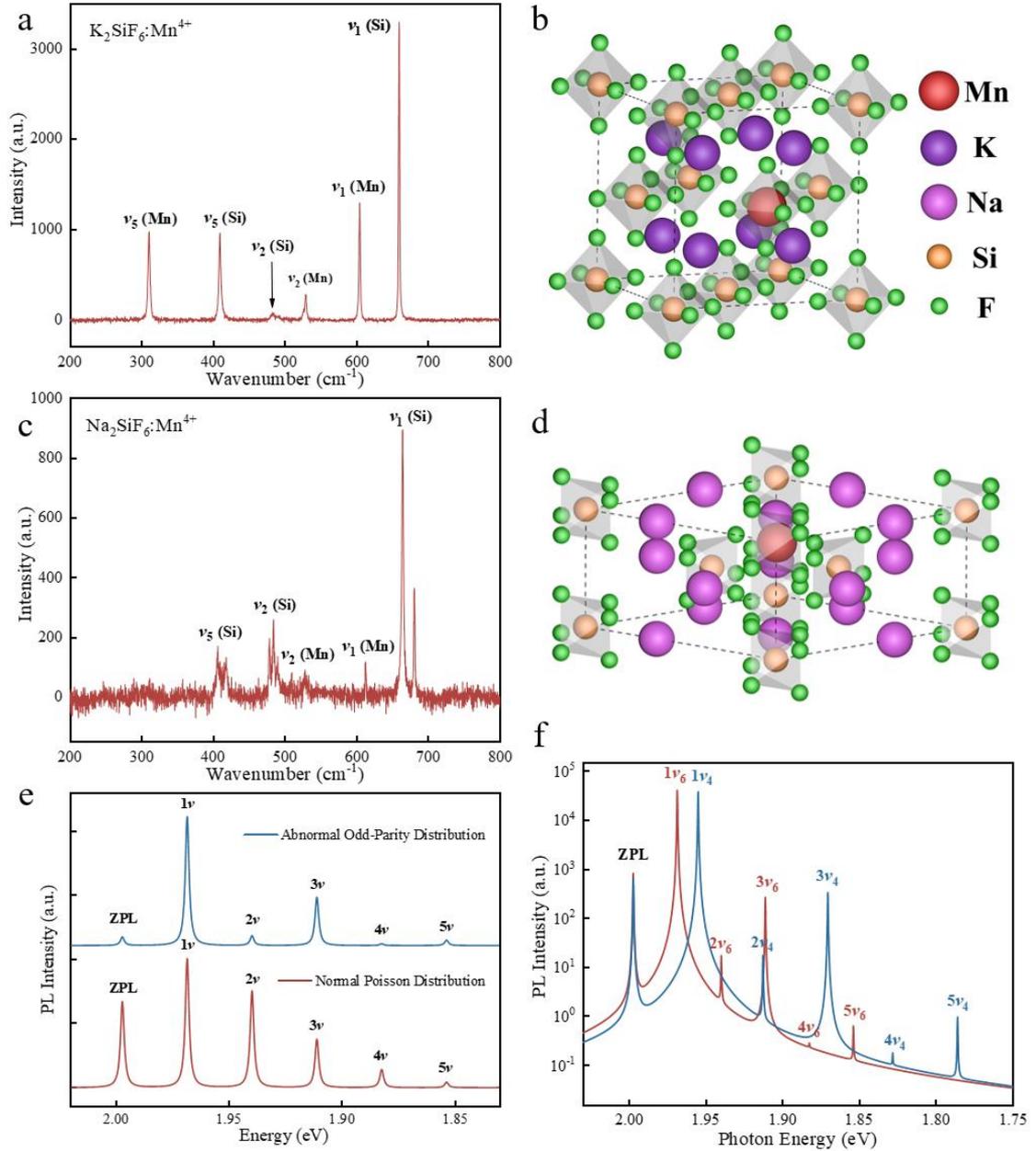

**Figure 1.** (a-d) Cryogenic Raman spectra and crystal structures of the studied $K_2SiF_6$:$Mn^{4+}$ and $Na_2SiF_6$:$Mn^{4+}$ phosphors. **(a)** Raman spectrum of the KSF phosphor measured at 8 K. **(b)** Crystal structure of the KSF phosphor (Space group: $Fm\bar{3}m$). **(c)** Raman spectrum of the NSF phosphor measured at 8 K. **(d)** Crystal structure of the NSF phosphor (Space group: $P321$). **(e)** Comparison of phonon sideband intensity distributions in parity-forbidden (top blue curve) versus conventional (bottom red curve) electron systems. **(f)** Schematic of parity-dependent phonon sidebands distributions of two phonon modes $v_4$ and $v_6$ in the KSF phosphor.

Figure 2 presents the semi-logarithmic high-resolution PL spectrum of the KSF phosphor measured at 8 K under the excitation of 405 nm laser. The prominent 1st-order phonon sidebands in the PL spectrum originate from the coupling between $Mn^{4+}$ 3d$^3$ electrons and asymmetric vibrational modes ($v_3$: asymmetric stretching mode; $v_4$, $v_6$: asymmetric bending mode), as observed in literature.[4-6,8,11,17] The ZPL and sidebands of $v_3$, $v_4$, and $v_6$ vibrational modes are explicitly labeled in the figure. More excitingly, the 3rd- and even 5th-order phonon sidebands of these modes can be resolved, as marked in the figure. The remaining weak peaks correspond to the acoustic phonons and combined modes, which fall beyond the scope of this investigation. A more complete deconvolution of the main spectral peaks of the $v_4$ and $v_6$ modes with Lorentzian line shape function is presented in Figure 1f. Notably, the observation of odd-higher-order phonon sidebands and the simultaneous suppression of the even-order phonon sidebands (marked in Figure 2 with Ø) in the KSF phosphor severely deviates from the Poisson intensity distribution predicted by the classic theories. This unravels an interesting phonon sideband intensity distribution for parity-forbidden material systems, i.e., normally distributed odd-order intensities versus abnormally suppressed even-order intensities, as demonstrated in Figure 1e. It should be noted that the parity-forbidden ZPL line can intrinsically corresponds to a 0th-order even-parity transition.

It is known that the phonon sideband peak intensities approximately satisfy a Poisson distribution in the classic theories where the Huang-Rhys factor $S$ can be approximated by the intensity ratio between the 1st-order sideband and the ZPL ($S \approx I_1/I_{ZPL}$).[13] Nevertheless, in parity-forbidden systems, the significant suppression of the even-order sidebands is not expected in the classic theories. If the 1st- and 3rd-order phonon sidebands retain Poisson statistics, the Huang-Rhys factor $S$ can be expressed as: $S = \sqrt{6I_3/I_1}$. Applying this formula to the fluorescence spectrum of the KSF phosphor, we yield $S$ factors of 0.017, 0.025, and 0.047 for the $v_3$, $v_4$, and $v_6$ vibrational modes, respectively. All $S$ values are substantially less than 1, revealing the existence of an extremely weak electron-phonon coupling in the studied KSF phosphors. This conclusion is in excellent agreement with the super high quantum efficiency of the KSF phosphors.

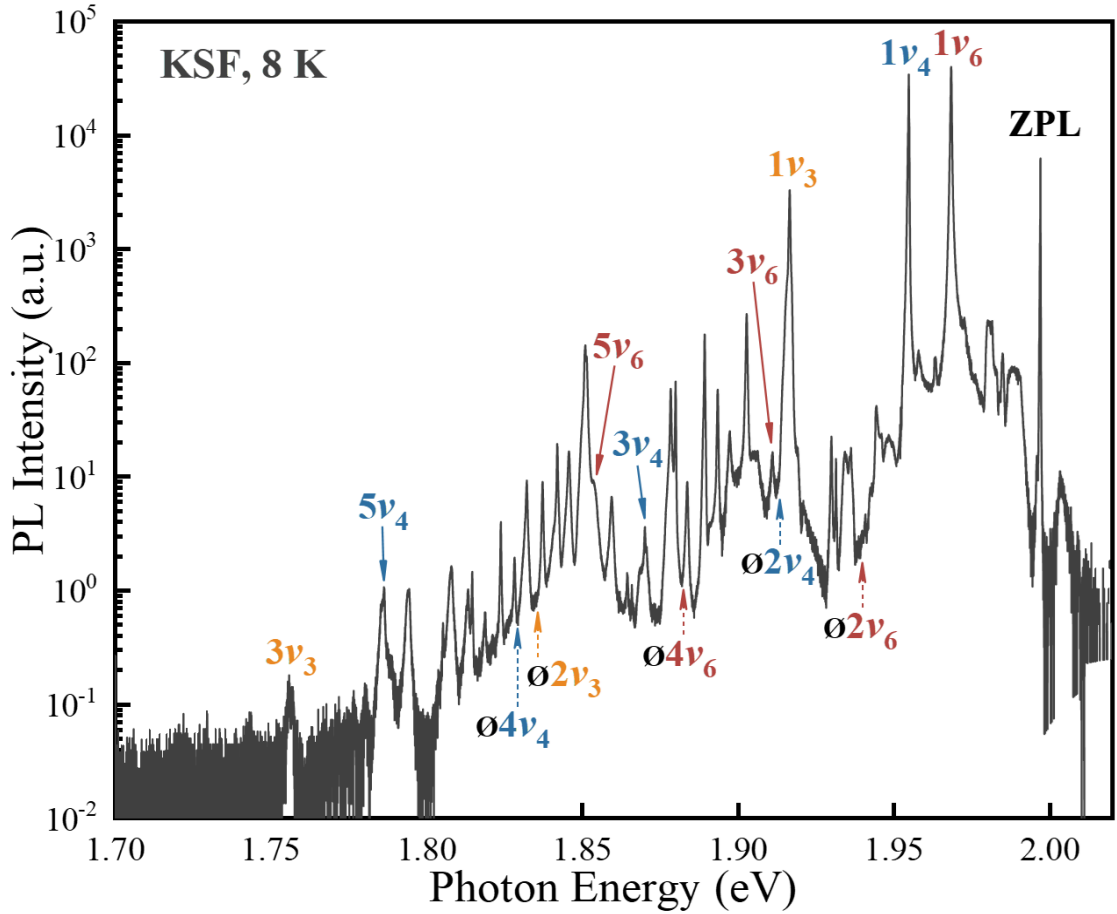

**Figure 2.** High-resolution PL spectrum (semi-logarithmic) of the KSF phosphor measured at 8 K. The predominate peaks are the 1st-order phonon sidebands of the three asymmetric vibration modes $v_3$, $v_4$, and $v_6$. In addition, the 3rd- and even 5th-phonon sidebands of these modes can be resolved, while the even-order phonon sidebands are strongly suppressed.

As a supplementary investigation, we also measured the micro-Raman spectrum (Figure 1c) of the NSF phosphor with trigonal crystalline structure (Figure 1d) and its high-resolution PL spectrum at 8 K (Figure 3). Since the NSF phosphor adopts a trigonal crystal structure, the $O_h$ symmetry of $[MnF_6]^{2-}$ octahedra is broken by the induced lattice distortion.[17,21] Such distortion is directly reflected in the Raman spectrum through pronounced peak splitting of $[SiF_6]^{2-}$-assigned modes, as seen in Figure 1c. Furthermore, the octahedral distortion and the resultant symmetry reduction are expected to significantly alter photophysical properties of the NSF phosphor, as unambiguously demonstrated in the PL spectrum in Figure 3. The PL spectral feature changes include:

1) Jahn-Teller Splitting of Energy Levels: Symmetry lowering induces splitting of the excited state ($^2E_g$), ground state ($^4A_{2g}$), and even the fundamental vibrational modes predicted by Jahn-Teller principle.[20] This is directly evidenced by the prominent multiple structures in the PL spectrum, featuring the labeled ZPL and 1st-order sidebands of the asymmetric vibrational modes $v_3$, $v_4$, and $v_6$.

2) Parity Selection Rule Relaxation and ZPL Enhancement: Reduced symmetry partially relaxes parity selection rules, leading to significant intensity enhancement of the originally parity-forbidden ZPL transition. The ZPL intensity in the NSF phosphor exhibits a significant enhancement relative to the ZPL line of the KSF phosphor—surpassing even the 1st-order phonon sideband intensity of the $v_6$ vibrational mode.

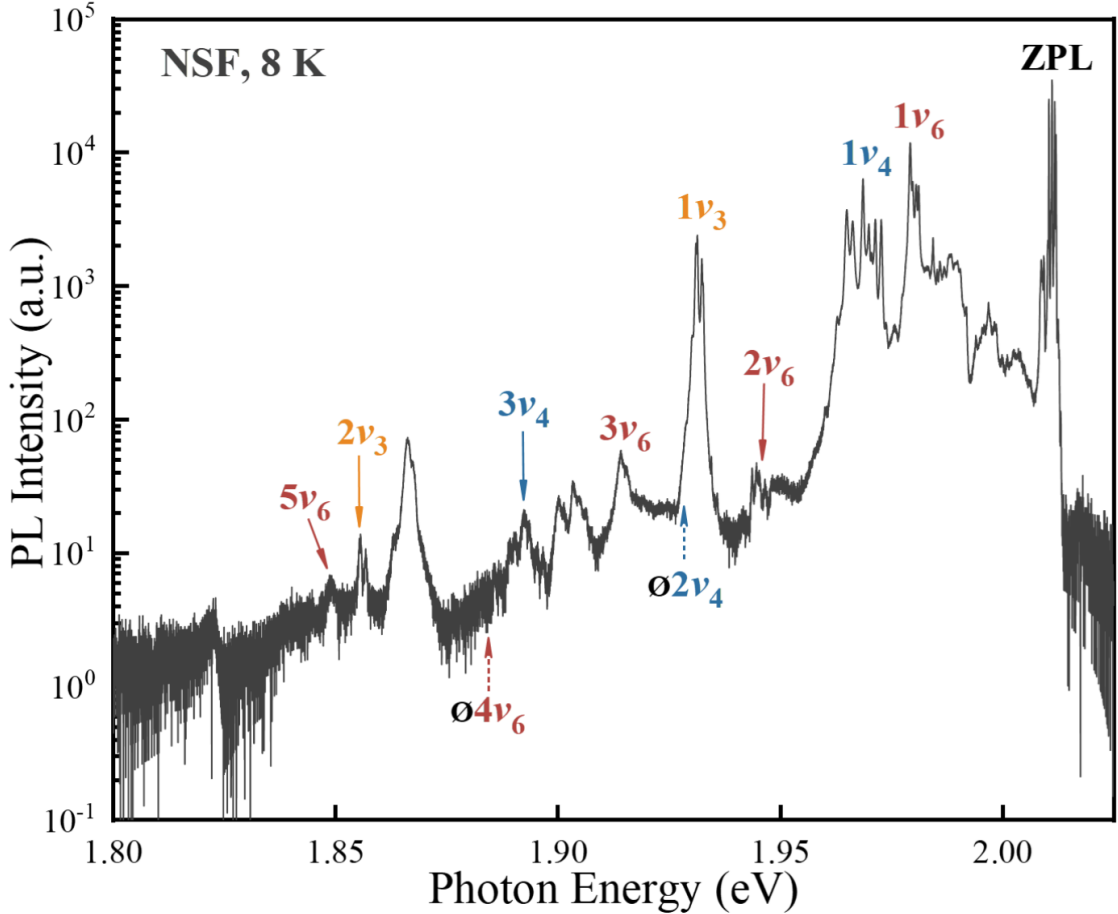

**Figure 3.** High-resolution PL spectrum (semi-logarithmic) of the NSF phosphor measured at 8 K. Compared with the PL spectrum of the KSF phosphor shown in Figure 2, the relative intensity of the ZPL line is significantly enhanced due to the parity selection rule relaxation induced by the symmetry lowering.

Consistent with the case of KSF, several odd-order phonon sidebands, such as $3v_6$ (3rd-order $v_6$ phonon sideband), $3v_4$, and even $5v_6$ can be resolved in the high-resolution PL spectrum. Despite that the ZPL line was significantly enhanced by the relaxed parity prohibition, its even-order sidebands (i.e., $2v_6$ and $2v_3$) were still very weak, as shown and marked in Figure 3. In order to identify the higher-order phonon sidebands more clearly, we implemented a comparative analysis of spectral line shapes between the 1st- (red curves) and the higher-order phonon sidebands (blue curves), as illustrated in Figure 4. It can be seen that all the resolved higher-order sidebands exhibit almost identical spectral features of their 1st-order sidebands, confirming the assignment of the higher-order phonon sidebands. Crucially, the PL spectrum of the NSF phosphor with trigonal structure and the significantly enhanced ZPL transition still show the odd-parity selection rule for the phonon-assisted electronic transitions. As will be demonstrated in subsequent derivations of modified transition matrix, this abnormal odd-parity phenomenon constitutes a direct consequence of partial lifting of parity selection rules.

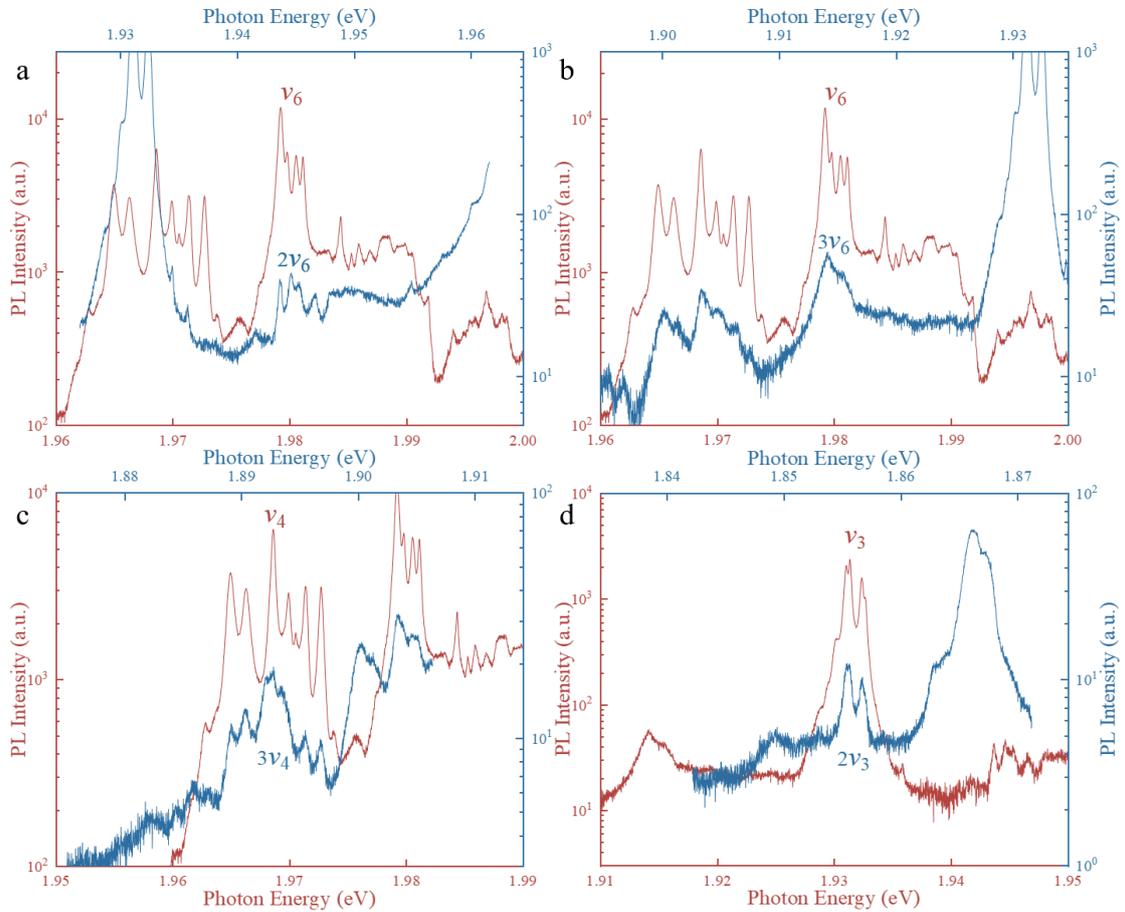

**Figure 4.** Comparison of the 1st-order sidebands (red curves) for $v_6$, $v_4$, $v_3$ with their higher-order phonon replicas (blue curves): (a) $2v_6$, (b) $3v_6$, (c) $3v_4$, (d) $2v_3$.

## 3 Theoretical Derivations and Discussions

The transition matrix element is the key parameter determining the emission intensity in PL spectra, whose amplitude can be expressed as $\langle i|\hat{M}|f\rangle$, where $|i\rangle$ and $|f\rangle$ represent the electronic wave functions of initial and final states in the optical transition, respectively, and $\hat{M}$ denotes the electric dipole moment operator. In electron-phonon coupling systems, the Condon approximation is widely employed, which factorizes the transition matrix element into a product of electronic transition and lattice vibration transition matrix elements:[16]

$$\langle i|\hat{M}|f\rangle \rightarrow \langle \mu'|e\hat{x}|\mu''\rangle \int X^*_{\mu'n'}(X)X^*_{\mu''n''}(X)dX, \tag{1}$$

where $\langle \mu'|e\hat{x}|\mu''\rangle$ represents the transition matrix element between two electronic states of the $F$-center, and X denotes the lattice vibrational wave function. In the $Mn^{4+}$-activated phosphors, the electronic transition between the initial ($^2E_g$) and final ($^4A_{2g}$) states of the d-orbital states is a parity-forbidden transition with a vanishing electric dipole transition matrix element. Under such circumstances, a first-order correction based on the Herzberg-Teller expansion has to be incorporated into the Condon approximation:[22]

$$\langle i|\hat{M}|f\rangle \rightarrow M_e(R_0)\langle X^*_{\mu'n'}|X^*_{\mu''n''}\rangle + \sum_k \left(\frac{\partial M_e}{\partial Q_k}\right)_{Q_k=0} \langle X^*_{\mu'n'}|Q_k|X^*_{\mu''n''}\rangle, \tag{2}$$

where $M_e(R) = \langle \mu'|e\hat{x}|\mu''\rangle$, and $Q_k$ denotes the coordinate of the $k$-th normal vibrational mode.

Following the derivation of Huang-Rhys theory, the lattice vibrational wave function X can be expressed as a product of normal oscillator wave functions $\prod_j X_{n'_j}(q'_j)$, where $q'_j = q_j - \frac{1}{\sqrt{N}}\frac{A'_j}{\omega_l^2}$ represents the modified normal coordinate, and $n_j'$ denotes the initial vibrational quantum number of the $j$-th normal oscillator.[12] The oscillator wave functions can be expanded in terms of $q_j'$, yielding

$$\int X^*_{\mu'n'}Q_k X^*_{\mu''n''}dX = \int \left\{Q_k \times \prod_j \left[X_{n'_j}(q_j) - \frac{1}{\sqrt{N}}\frac{A'_j}{\omega_l^2}X'_{n'_j}(q_j) + \frac{1}{2}\left(\frac{1}{\sqrt{N}}\frac{A'_j}{\omega_l^2}\right)^2 X''_{n'_j}(q_j) + \cdots\right]\right.$$
$$\left. \times \left[X_{n''_j}(q_j) - \frac{1}{\sqrt{N}}\frac{A''_j}{\omega_l^2}X'_{n''_j}(q_j) + \frac{1}{2}\left(\frac{1}{\sqrt{N}}\frac{A''_j}{\omega_l^2}\right)^2 X''_{n''_j}(q_j) + \cdots\right]\right\}\prod_j dq_j. \tag{3}$$

Based on the orthogonality and recurrence relations of Hermite polynomials, all transitions involving changes in vibrational quantum numbers greater than 1 can be neglected.[23] Under the Huang-Rhys theoretical framework, we consider a system comprising $j$ harmonic oscillators, where $s$ oscillators go down by one quantum and $s+p$ oscillators go up by one quantum.[12] Using indices $l$ and $k$ to label these two types of oscillators, respectively, we multiply the series in Eq. (3) and then neglect second-order and higher terms to obtain:

$$\int X^*_{\mu'n'} Q_k X^*_{\mu''n''} dX = \int \left\{ Q_k \times \prod_j \left[ X^2_{n'_j}(q_j) - \frac{2}{\sqrt{N}} \frac{A'_j}{\omega_j^2} X_{n'_j}(q_j) X'_{n'_j}(q_j) \right] \right.$$

$$\times \frac{\prod_l \left[ X_{n'_l}(q_l) X_{n'_l+1}(q_l) - \frac{1}{\sqrt{N}} \frac{A''_l}{\omega_l^2} X_{n'_l}(q_l) X'_{n'_l+1}(q_l) - \frac{1}{\sqrt{N}} \frac{A'_l}{\omega_l^2} X_{n'_l+1}(q_l) X'_{n'_l}(q_l) \right]}{\prod_l \left[ X^2_{n'_l}(q_l) - \frac{2}{\sqrt{N}} \frac{A'_l}{\omega_l^2} X_{n'_l}(q_l) X'_{n'_l}(q_l) \right]}$$

$$\times \frac{\prod_k \left[ X_{n'_k}(q_k) X_{n'_k-1}(q_k) - \frac{1}{\sqrt{N}} \frac{A''_k}{\omega_l^2} X_{n'_k}(q_k) X'_{n'_k-1}(q_k) - \frac{1}{\sqrt{N}} \frac{A'_k}{\omega_l^2} X_{n'_k-1}(q_k) X'_{n'_k}(q_k) \right]}{\prod_k \left[ X^2_{n'_k}(q_k) - \frac{2}{\sqrt{N}} \frac{A'_k}{\omega_l^2} X_{n'_k}(q_k) X'_{n'_k}(q_k) \right]} \left. \right\} \prod dq . \quad (4)$$

Considering the parity properties of each term in the expression, the normal coordinate $Q_k$ possesses odd parity. The parity of harmonic oscillator eigenfunctions is determined by the quantum number $n$: The wavefunction exhibits odd parity when $n$ is odd, and even parity when $n$ is even.[24] The differential operator $\partial/\partial q_k$ acts as a parity-flipping operator. Table 1 summarizes the parity characteristics of various terms in the electron-phonon coupling expansion. According to the parity selection rule, the transition matrix element integral yields a non-zero value only when the complete integrand maintains even parity.

Table 1   Relationship between parity characteristics and order

| $\Delta n_j$ | 0th-order component | 1st-order component |
| --- | --- | --- |
| 0 | Even | Odd |
| 1 | Odd | Even |
| (2) | Even | Odd |

The product of all the 0th-order terms, which dominates the matrix element contribution, consists of ($j$-2$s$-$p$) even-parity factors with unchanged quantum numbers, $s$ odd-parity factors go down by one quantum, and ($s$+$p$) odd-parity factors go up by one quantum. The net parity of this 0th-order term is given by the product $odd \times (even)^{j-2s-p} \times (odd)^{s+p} \times (odd)^s = (odd)^{p+1}$, demonstrating that only when $p$ is odd, the terms possess overall even parity and make non-zero contributions to the transition matrix element. For even $p$ values, the 0th-order term vanishes due to odd net parity, and the 1st-order term becomes the primary contributor, although its magnitude remains substantially smaller than the potential 0th-order contributions. Since $p$ corresponds directly to the phonon sideband order, this parity selection rule leads to markedly stronger intensities for the odd-order phonon sidebands compared to their even-order counterparts in parity-forbidden electronic transition systems (e.g., the studied KSF phosphor).

For parity-partially-relaxed systems (e.g., the studied NSF phosphor), partial lifting of parity restrictions restores contributions from the Condon approximation terms, providing a baseline intensity for all orders sideband orders. Simultaneously, the non-zero contributions from the Herzberg-Teller term persist. Since the Herzberg-Teller term enhances only the odd-order phonon sidebands and the even-order sidebands rely solely on the Condon term, the interplay of these two mechanisms results in a characteristic spectral pattern in which the odd-order sidebands exhibit systematically higher intensities than their adjacent even-order counterparts, as clearly observed in Figure 3.

Subsequent derivations will establish explicit mathematical forms for the odd-order phonon sideband intensities. Under the assumption of uniform response of the electronic transition matrix element to all normal modes, Eq. (2) reduces to the simplified form:

$$\langle i|\widehat{M}|f\rangle \to M_e(R_0)\langle X^*_{\mu'n'}|X^*_{\mu''n''}\rangle + \left(\frac{\partial M_e}{\partial Q}\right)_{Q=0} \sum_k \langle X^*_{\mu'n'}|Q_k|X^*_{\mu''n''}\rangle. \qquad (5)$$

Here, $Q_k$ denotes the normal coordinate of the $k$-th normal vibrational mode. If we assume that $\bar{n}$ represents the thermal average of the vibrational quantum number and is identical for all oscillators, and denoting the oscillator index coupled to $Q_k$ as $i$, the total transition matrix element partitions into three components:

$$\langle i|\widehat{M}|f\rangle \to \left(\frac{\partial M_e}{\partial Q}\right)_{Q=0} (M_1 + M_2 + M_3), \qquad (6)$$

where $M_1$, $M_2$, and $M_3$ represent the contributions from normal vibrational modes with the quantum number unchanged, going up by one, and down by one, respectively, which are given by:

$$M_1 = (j - 2s - p)\langle \bar{n}|Q_i|\bar{n}\rangle \prod_{j \neq i} \int X_{n'_j}(q'_j) X_{n'_j}(q''_j) dq_j \times \prod_l \frac{\int X_{n'_l}(q_l) X_{n'_l+1}(q_l) dq_l}{\int X_{n'_l}(q'_l) X_{n'_l}(q''_l) dq_l}$$

$$\times \prod_k \frac{\int X_{n'_k}(q_k) X_{n'_k-1}(q_k) dq_k}{\int X_{n'_k}(q'_k) X_{n'_k}(q''_k) dq_k}, \qquad (7)$$

$$M_2 = (s + p)\langle \bar{n}|Q_i|\bar{n}+1\rangle \prod_j \int X_{n'_j}(q'_j) X_{n'_j}(q''_j) dq_j \times \prod_{l \neq i} \frac{\int X_{n'_l}(q_l) X_{n'_l+1}(q_l) dq_l}{\int X_{n'_l}(q'_l) X_{n'_l}(q''_l) dq_l}$$

$$\times \prod_k \frac{\int X_{n'_k}(q_k) X_{n'_k-1}(q_k) dq_k}{\int X_{n'_k}(q'_k) X_{n'_k}(q''_k) dq_k}, \tag{8}$$

and

$$M_3 = s\langle \bar{n}|Q_i|\bar{n}-1\rangle \prod_j \int X_{n'_j}(q'_j) X_{n'_j}(q''_j) dq_j \times \prod_l \frac{\int X_{n'_l}(q_l) X_{n'_l+1}(q_l) dq_l}{\int X_{n'_l}(q'_l) X_{n'_l}(q''_l) dq_l}$$

$$\times \prod_{k\neq i} \frac{\int X_{n'_k}(q_k) X_{n'_k-1}(q_k) dq_k}{\int X_{n'_k}(q'_k) X_{n'_k}(q''_k) dq_k}. \tag{9}$$

Invoking the operational properties of harmonic oscillator wavefunctions and fundamental conclusions from the Huang-Rhys theory, the transition matrix element for odd $p$ can be modified as:

$$\langle i|\hat{M}|f\rangle \to \left(\frac{\partial M_e}{\partial Q}\right)_{Q=0} \sum_{s=0}^{\infty} \left[(s+p)\sqrt{\frac{\bar{n} S_+^{s+p-1} S_-^s}{s!(s+p-1)!}} + s\sqrt{\frac{(\bar{n}+1)S_+^{s+p} S_-^{s-1}}{(s-1)!(s+p)!}}\right] e^{-\frac{S_+ + S_-}{2}}, \tag{10}$$

where $S_\pm$ are defined as in Ref. [12]. Considering a system with weak electron-phonon coupling (e.g., the fluoride phosphors investigated in this study), where the Huang-Rhys factor $S$ and $S_\pm$ both should be much less than 1, the summation term in the formula exhibits a rapid decay with increasing quantum number $s$. Therefore, higher-order terms with $s \geq 1$ can be neglected, leading to an approximate expression for the intensity of odd-order phonon sidebands as:

$$I(p) \to \nu\bar{n}\left(\frac{\partial M_e}{\partial Q}\right)_{Q=0}^2 \frac{p^2 S_+^{p-1}}{(p-1)!} e^{-(S_+ + S_-)}. \tag{11}$$

In above equation, $\nu$ stands for the frequency of emitted photons. When the temperature is extremely low (i.e., $k_B T \ll \hbar\omega$), the thermal average vibrational quantum number $\bar{n}$ is much smaller than 1. If the energy differences among the various phonon sidebands are further neglected, Eq. (11) can be simplified to:

$$I(p) \propto \frac{p^2 S^{p-1}}{(p-1)!} e^{-S}. \tag{12}$$

Compared with the Poisson distribution derived from the classical Huang-Rhys theory, Eq. (12) shows an abnormal phonon sideband intensity distribution of the odd-order phonon sidebands of the parity-forbidden electronic transitions. Based on the above formula, the Huang-Rhys factor $S$ can be expressed as: $S = \sqrt{2I_3/9I_1}$, which diverges markedly from the value of $S$ previously derived from the Poisson distribution. The newly derived $S$ factor is approximately one-fifth of the original value in the Poisson distribution. For the KSF phosphors investigated in the present study, their *S* value was on the order of $10^{-3}$, indicating that the material system indeed belongs to the catalog of extremely weak coupling regime.[13] Such weak but distinct electron-phonon coupling endows the outstanding luminescence properties of $Mn^{4+}$-activated fluoride phosphors.

## 4 Conclusions

In summary, the spectral signatures of the 3rd- and 5th-order phonon sidebands were firmly resolved in the high-resolution fluorescence spectroscopy of the $Mn^{4+}$-doped fluoride phosphors, while the even-order phonon sidebands were almost unobserved. This pronounced odd-even intensity alternation directly challenges the existing theoretical understanding about the phonon-assisted PL in solid luminescent materials. By introducing the Herzberg-Teller approximation as the first-order correction to the Condon approximation in the treatment of transition matrix elements, the physical origin of this phenomenon is theoretically unraveled: The parity selection rules strongly suppress the even-order sidebands through inhibiting the 0th-order transition terms in parity-forbidden systems. The self-consistent agreement between experiment and theory not only resolves the long-standing puzzling about the absence of the higher-order phonon sidebands in such systems, but also provides a fundamental framework for understanding the electron-phonon coupling mechanisms in parity-forbidden electronic transition systems.

Acknowledgments: The work was financially supported by the National Natural Science Foundation of China (No. 12074324). The authors wish to thank Dr. Debao Zhang, Dr. Ji Zhou, Mr. Wanggui Ye, and Mr. Xuguang Cao for their important contributions to the construction of the multi-function-integrated micro-spectroscopy system and contributions to the work.